\documentclass[journal]{IEEEtran}
\usepackage{cite}
\usepackage{graphicx}
\usepackage{booktabs}
\usepackage[table]{xcolor}

\usepackage{url}
\usepackage{epstopdf}
\usepackage{amsfonts,amsmath,amsthm,amssymb}
\usepackage{subfigure}
\usepackage{stfloats}
\usepackage{eurosym}
\usepackage{enumerate}
\usepackage{color}
\usepackage{arydshln}
\usepackage{longtable}
\usepackage{mathrsfs}
\usepackage{bm}
\usepackage[T1]{fontenc}
\usepackage[latin1]{inputenc}
\usepackage[shortlabels]{enumitem}
\usepackage{tikz}
\usepackage{footnote}
\makesavenoteenv{minipage}
\makesavenoteenv{enumerate}
\usepackage{endnotes}
\usetikzlibrary{arrows}

\usepackage{csquotes}

\usepackage{psfrag}

\usepackage{balance}

\newtheorem*{remark}{Remark}

\newcommand{\be}{\begin{equation}}
\newcommand{\ee}{\end{equation}}
\newcommand{\bea}{\begin{eqnarray}}
\newcommand{\eea}{\end{eqnarray}}

\newcommand{{\bgamma}}{{\Vector \gamma}}

\def\pro2{\tilde {\mathcal P}}

\def\Xi{X^{(i)}}

\def\<{\left< }
\def\>{ \right>}
\def\t>{ \right>_{T_w}}

\def\tt{\tilde t}

\def\Vector#1{\mbox{\boldmath $#1$}}

\def\1{{\rm 1}}

\def\s1{^{\rm (1)}}

\begin{document}

\graphicspath{{Figures/}}

\title{Optimal partitioning of multi-thermal zone buildings for decentralized control}
\author{Ercan Atam,
       Eric C. Kerrigan
\thanks{
Ercan Atam and Eric C. Kerrigan are with Electrical and Electronic Engineering Department of
Imperial College London, South Kensington Campus, SW7 2AZ, London, United Kingdom. Kerrigan is also with the Department of Aeronautics.
E-mails: e.atam@imperial.ac,uk, e.kerrigan@imperial.ac,uk.}}


\maketitle

\begin{abstract}

In this paper, we develop an optimization-based systematic approach for the challenging, less studied, and important problem of optimal partitioning of multi-thermal zone buildings for the decentralized control.
The proposed method consists of
(i) construction of a graph-based network to quantitatively characterize the thermal interaction
level between neighbor zones, and (ii) the application of two different approaches for optimal clustering of
the resulting network graph: stochastic optimization and robust optimization.
The proposed method was tested
on two case studies: a 5-zone building (a small-scale example) which allows one to consider all possible partitions
to assess the success rate of the developed method; and a 20-zone building (a large-scale example) for which the developed method was used to predict the optimal partitioning of the thermal zones.
Compared to the existing literature, our approach provides a systematic and potentially optimal
solution for the considered problem.
\end{abstract}

\IEEEpeerreviewmaketitle

\section{Introduction}

Among  building thermal comfort control strategies, model predictive control (MPC)
and its variants are the most popular ones \cite{Oldewurtel_et_al_2012, Oldewurtel_et_al_2014, Atam_2016, Atam_2017},
which can provide energy savings of up to 40\%  compared to an on-off or rule-based controller \cite{Privara_et_al_2011,
Sturzenegger_et_al_2013_conf,Dong_Lam_2014,Bengea_et_al_2014}.
In energy-efficient thermal control of buildings, both centralized MPC (C-MPC) and decentralized MPC (D-MPC)
can be used. Given a multi-zone building consisting of a large number of thermal zones (such as a university building or
an airport), it is crucial to have a kind of optimal balance between centralization and decentralization \cite{Atam_2016_dec_paper},
which in turn means that one needs to find an optimal partitioning of thermal zones into a set of clusters and control of
each cluster using a separate C-MPC. In optimal partitioning  the objective is to partition the
building thermal zones into a set of clusters to achieve (i)~minimum thermal coupling between the clusters, (ii)~strong
thermal connectivity between members of each cluster, (iii)~numerically-efficient
control of each cluster, (iv)~acceptable performance loss compared to C-MPC when the D-MPC is applied to clusters.

Multi-thermal zone partitioning approaches can be divided into
two categories: approaches where control design is not an integrated
part of the partitioning approach and  \textit{co-design-based partitioning}  where control design is an integrated part of the partitioning. Although the problem of optimal partitioning of multi-zone buildings for
decentralized control is a very important problem to be solved, there have appeared  very few studies in the
literature on this topic, probably due to the challenges involved.

We start the literature review by mentioning three existing works which use approaches in the first category. In \cite{Cai_Braun_2014},
they modelled the interior wall between two neighbor thermal
zones  either using the 3R2C structure if the wall has a high thermal mass, or
1R structure if the wall has significant openings. To quantify the level of thermal interaction
between a given zone and a neighbor zone, the middle resistance in 3R2C or the single
resistance in 1R was set to a nominal value and then to a
large value (which basically means full de-coupling). Next, by finding the
temperature difference between the two cases and normalizing the difference,
they obtained a measure of the thermal interaction. Repeating this for all
neighboring zones, they quantified inter-zone thermal interactions of a multi-zone building.
Next, by specifying a threshold, they eliminated the couplings where the interaction degree is below the threshold,
which in  turn resulted in a ``manual" partitioning of the zones into a ``non-predetermined"
number of clusters where medium-to-high
thermal interactions exist between members in each cluster.
In \cite{Agbi_2014}, the concepts of graph theory-based structural and
output identifiability \cite{Doren_et_al_2009, Doren_et_al_2011} together with the relevant metrics were used
to decompose a multi-zone building into identifiable clusters satisfying both
structural and output identifiability. Next, using decentralized uncertain models,
a two-level hierarchical robust MPC
scheme was developed for control of the whole building system.
Although the schemes for both identification and control
were regarded as ``decentralized", the thermal interactions between a zone and
the neighbor zones were explicitly taken into account as dynamic interaction signals using
the outputs of models of these zones. As a result, the term ``decentralization" in this work is not used
in the same meaning of our definition which is no use of such a dynamic interaction signal,
and  hence the work of \cite{Agbi_2014} can be
seen more as a kind of distributed identification
and control. The last work in the first category 
is the work done in \cite{Kyriacou_et_al_2017} for the related problem of
contaminant detection and isolation in multi-zone buildings. They presented
two methods. (i) An optimization-based graph clustering method
where mixed-integer linear programming was used to formulate the
optimization problem and the objective was optimal decentralization of the system to minimize interconnection airflows between clusters.
(ii) A heuristic method based on matrix clustering to decentralize the system where the
developed heuristic is computationally faster than the optimization-based graph clustering approach
but suboptimal compared to it.

As an example of co-design-based thermal zone partitioning  studies we can mention only one study. In \cite{Chandan_et_al_2013},
an analytically-derived optimality loss factor was used to  measure the
optimality difference between an unconstrained decentralized control architecture and unconstrained centralized MPC. Next, the developed
optimality loss factor was used as a distance metric for optimal partitioning based on an agglomerative clustering approach. The main
limitation of \cite{Chandan_et_al_2013} is that no constraint
was taken into account in the MPCs.

In this paper, we present a new method (which can be put
into the first category of zone partitioning categories mentioned before) for zone partitioning and our contributions can be summarized as follows.
(i) First, we build on an idea similar to the thermal interaction-quantification
idea presented in \cite{Cai_Braun_2014} to determine ``interval"-type thermal interaction levels
between zones, which is more realistic compared to ``single"-type thermal interaction levels of \cite{Cai_Braun_2014}.
(ii) We formulate the optimal clustering of interval-weighted thermal interaction network graph using
a stochastic optimization and a robust optimization framework.
(iii) We define a new index for performance of any control architecture, then using this index
we defined an ``optimality deterioration" and a ``fault propagation" metric to assess the performance
of C-MPC/D-MPC architectures.
(iv) We demonstrated the effectiveness of the presented approach by testing the developed algorithm on two case studies.

The rest of the paper is organized as follows. In Section
\ref{sec:Formulation of open-loop-based optimal partitioning approach}
the details of graph-based optimization formulation (including both robust
and stochastic frameworks) of the developed partitioning approach are given.
Next, in this section we define a performance index to measure
the optimality of any control method for thermal comfort of buildings, and then
using this index we define the optimality deterioration, fault propagation metrics
and a combined metric to determine the best partition
among the best partitions. Two case studies are given in Section \ref{sec:Case studies} to demonstrate
the developed approach and the results are discussed in detail. Finally, 
the conclusions and some future research work are given in Section \ref{sec:Conclusions}.

\section{Formulation of optimal partitioning approach}
\label{sec:Formulation of open-loop-based optimal partitioning approach}


\subsection{Building thermal model}
\label{subsec:Building thermal model}

The building thermal dynamics models were developed in the BRCM toolbox \cite{Sturzenegger_et_al_2014}
which is a high-fidelity toolbox where the thermal models are developed following a thermal resistance-capacitance network
approach. The models of the toolbox were validated against EnergyPlus \cite{EnergyPlus} (a popular building thermal dynamics and energy simulation platform) and it was observed that the temperature difference between the two software is  less than 0.5 \textdegree{}C
\cite{Sturzenegger_et_al_2014}.
The models have the form given in \eqref{emulator_model_LSMZB}
\begin{subequations} \label{emulator_model_LSMZB}
\begin{align}
x(k+1)=& Ax(k)+B_uu(k)+B_ww(k),\\
T(k)=& Cx(k).
\end{align}
\end{subequations}
\noindent with $u$ denoting the control
input (heating or cooling power), $w$ the
external predictable disturbances (ambient air
temperature, solar radiation and internal gains inside the building)
and $T$ the zone air temperatures. The sampling time was set to 15 minutes.

Let the multi-zone building be represented as the undirected connected graph $G\triangleq(V, E)$
where $V\triangleq\{1, 2, \ldots, N_z \}$ denotes the set of indexed $N_z$ thermal zones  and $E$
denotes the weighted edges, where the edge weights are intervals representing the level of thermal interaction between neighbor zones.
We assume that the building zones can be partitioned into $n$ clusters, where $ 1 \le n \le N_z$.

\subsection{Thermal interaction quantification and the thermal interaction graph}
\label{subsec:Thermal interaction quantification and the thermal interaction graph}

In this section, we  present a formula for calculation of
thermal interaction intervals between neighboring thermal zones.
Let $w$ denote the disturbance vector, which consists of internal gains, ambient temperature
and solar radiation data over a representative year; let $T_{cr} \triangleq [T_l(k), T_u(k)]$ be the comfort range
for occupants (which can be time-dependent). Next, we define a modified comfort range
$\tilde{T}_{cr} \triangleq [T_l(k)-1, T_u(k)+1]$  obtained from allowing
1\textdegree{}C temperature violation which can happen due to model or disturbance
uncertainties during control design and its implementation in real control of buildings. 
Now we will construct a simulation input, $\mathcal{U}_{\text{ti}}$, as follows:
based on $\tilde{T}_{cr}$, we construct
random zone temperature set-points lying in $\tilde{T}_{cr}$ (that characterize
possible controlled zone temperatures) and then we obtain the corresponding control
input vector $u$ which will track these set-points under the influence of $w$.
We take $\mathcal{U}_{\text{ti}}$ as $(u,w)$
where $u$ is random but consistent with building  control inputs encountered during controlled building operation
to produce zone temperatures that stay in $\tilde{T}_{cr}$.
It is very important that $\mathcal{U}_{\text{ti}}$ includes disturbances and control inputs
of controlled building during different seasons, different periods in a day (day time and night time),
and different weather conditions. Hence, it is crucial to consider a one-year period using a representative year
when $\mathcal{U}_{\text{ti}}$ is constructed.
Now, let $T_i(u,w)$ be the temperature of the $i$-th zone obtained
from the response of thermal model $M$ when it is simulated with input $(u,w)$ and
let $T_i^{\text{ no}\,j}(u,w)$ be the corresponding response when the dynamics of the
$j$-th neighbor zone is decoupled from $M$. The interval including the
thermal interaction degrees between the neighbor $i$-th and $j$-th zone is then given
by
\begin{subequations} \label{thermal_interaction_interval_comp}
\begin{align}
\underbar{I}_{ij}^{'}=&\min_{k} \left|T_i\big(u(k),w(k)\big)-T_i^{\text{ no}\,j}\big(u(k),w(k)\big)\right|,\\
\bar{I}_{ij}^{'}=&\max_{k} \left|T_i\big(u(k),w(k)\big)-T_i^{\text{ no}\,j}\big(u(k),w(k)\big)\right|,\\
\underbar{I}_{ij}=&\frac{\underbar{I}_{ij}^{'}+\underbar{I}_{ji}^{'}}{2},\label{thermal_interaction_interval_comp_equ_c} \\
\bar{I}_{ij}=&\frac{\bar{I}_{ij}^{'}+\bar{I}_{ji}^{'}}{2},\label{thermal_interaction_interval_comp_equ_d}\\
I_{ij}=& [\underbar{I}_{ij},\,\bar{I}_{ij}].
\end{align}
\end{subequations}
Here note that \eqref{thermal_interaction_interval_comp_equ_c}-\eqref{thermal_interaction_interval_comp_equ_d} are used
to obtain a single average thermal interaction interval between two
neighbor zones.

\begin{remark} [Design of $\mathcal{U}_{ti}$]
It is important that the system is in closed-loop operation (but not necessarily operated by an optimal controller)
since we are interested in inter-zone thermal interaction levels to be encountered
when building is under control. These are the interaction degrees which
determine the degradation level of a decentralized control architecture compared
to a centralized one.
\end{remark}

\subsection{Formulation of optimal partitioning problem}
\label{subsec:Formulation of optimal partitioning problem}

The optimization-based optimal partitioning formulation of multi-thermal
zone buildings consists of formulation of two types of constraints and the objective function.
We start with the first set of constraints known as ``Cluster formation constraints"
which are compactly given in \eqref{Cluster formation constraints}.
A subset of these constraints were also considered in partitioning formulation
of some other problems \cite{Boulle_2014, Shirabe_2005, Conrad_et_al_2007, Kyriacou_et_al_2017, Dilkina_Gomes_2010} different from the multi-zone partitioning problem which is considered in this study the first time  to the best of authors' knowledge.
We assume that the set of thermal zones are partitioned into $n$ clusters where
we represent a generic cluster by $c$ and the cluster set by $\mathcal{C}=\{1, 2, \ldots, n\}$.
Let $p_{i, c}$ denote a binary variable indicating whether in the $c$-th cluster
the $i$-th thermal zone is included ($p_{i, c}=1$) or not ($p_{i, c}=0$);
$s_{i, j, c}$ a binary variable indicating whether in the $c$-th cluster
the edge $(i,j)$ is included ($s_{i, j, c}=1$) or not ($s_{i, j, c}=0$); and
$r_{i,j}$ a binary variable indicating whether the edge $(i,j)\in E$
is included in any cluster ($r_{i,j}=1$) or not ($r_{i,j}=0$).
Note that when clusters are formed, a subset of edges may not lie in any of these clusters, but rather they
may be ``crossing edges" between clusters.

\begin{subequations} \label{Cluster formation constraints}
\begin{eqnarray}
&\displaystyle \sum_{c=1}^np_{i, c}=1,                          &i \in V,                         \label{cfbc_1} \\
&\displaystyle \sum_{i=1}^{N_z}p_{i, c} \ge 1,                  &c \in \mathcal{C},               \label{cfbc_2} \\
&\displaystyle \sum_{c=1}^ns_{i, j, c} \le 1,                   &(i,j) \in E,                     \label{cfbc_3} \\
&s_{i, j, c} \le p_{i, c},                                       &(i,j) \in E, c \in \mathcal{C},  \label{cfbc_4} \\
&s_{i, j, c} \le p_{j, c},                                       &(i,j) \in E, c \in \mathcal{C},  \label{cfbc_5} \\
&p_{i, c}+ p_{j, c} \le s_{i, j, c}+1,                          &(i,j) \in E, c \in \mathcal{C},  \label{cfbc_6} \\
&r_{i, j}=\displaystyle\sum_{c\in \mathcal{C}}s_{i,j,c},        &(i,j) \in E,                     \label{cfbc_7} \\
&p_{i, c} \in \{0,1\},                                          &c \in \mathcal{C}, i \in V.      \label{cfbc_8} \\
&s_{i, j, c} \in \{0,1\} \equiv s_{i, j, c} \in [0,1] ,         &c \in \mathcal{C}, (i,j) \in E,  \label{cfbc_9} \\
&r_{i, j} \in \{0,1\} \equiv r_{i, j} \in [0,1],                &(i,j) \in E,                     \label{cfbc_10}
\end{eqnarray}
\end{subequations}
In  cluster formation constraints, \eqref{cfbc_1} indicates that a thermal zone should
be included in only one cluster;
\eqref{cfbc_2} indicates that each cluster should include at least one zone;
\eqref{cfbc_3} indicates that an edge can be included in at most one cluster;
\eqref{cfbc_4}-\eqref{cfbc_5} indicate that for an edge to be
included in a cluster both of its vertices should be included; \eqref{cfbc_6} indicates
that if both $i$-th and $j$-th zones are included in cluster $c$, then the edge
$(i,j)$ should also be included (the connectivity constraint); \eqref{cfbc_7} is used to know 
if the  edge $(i,j)$ is included in any cluster or not;
and finally \eqref{cfbc_8}-\eqref{cfbc_10} indicate that the variables  $ p_{i, c}$, $s_{i, j, c}$ and
$r_{i, j}$ are binary variables, but among which $s_{i, j, c}$ and
$r_{i, j}$, equivalently, can be considered as continuous variables
in the interval $[0, 1]$ with the following reason: case 1:  $p_{i,c}=0 \text{ or } p_{j,c}=0$
implies $s_{i, j, c}=0$ by \eqref{cfbc_4}-\eqref{cfbc_5},
case 2:  $p_{i, c}=1 \text{ and } p_{j, c}=1$
implies $s_{i, j, c}=1$ by~\eqref{cfbc_6}, and the equivalence in $\eqref{cfbc_10}$
follows from \eqref{cfbc_7}. Note that if we want, we can eliminate $r_{i, j}$ since
it is a dependent variable, but we will keep it to make the presentation clear.

\begin{figure}[t!]
\centering
\includegraphics[width=0.5\columnwidth]{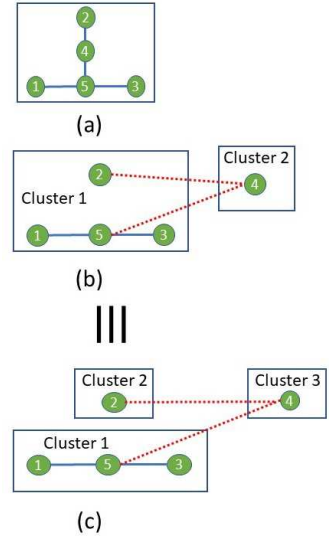}
\caption{An example to illustrate equivalence of disconnected clusters to  higher-order connected clusters: (a) the
thermal network, (b) disconnected two-cluster partition, (c) equivalent (in terms of decentralized control) connected three-cluster partition.}
\label{fig:connectivity_pic}
\end{figure}

\begin{remark}[Connectivity of clusters]
As illustrated in Figure \ref{fig:connectivity_pic}, in terms of D-MPC design,
a disconnected cluster is equivalent to a higher-order connected cluster, since for a disconnected
group in a cluster the associated model and hence the corresponding D-MPC has no coupling
with the rest. As a result, in multi-zone building partitioning problems, it is enough to consider
only connected clusters.
 \end{remark}

Next, we preset the size and relative size constraints in \eqref{Cluster size constraints} regarding
the obtained clusters. Such constraints are important for the management of
the computational load of cluster level control design and/or its sensitivity to any fault.
\begin{subequations} \label{Cluster size constraints}
\begin{eqnarray}
&\displaystyle \underbar{M}_a \le \displaystyle \sum_{i \in V}p_{i, c}  \le \bar{M}_a, \, c \in \mathcal{C},
\label{size_c_1} \\
&\displaystyle \left|\sum_{i \in V}p_{i, c_a}-\sum_{i \in V}p_{i, c_b}\right|  \le M_r, \, c_a, c_b \in \mathcal{C}, \,a \ne b. \, \label{size_c_2}
\end{eqnarray}
\end{subequations}
Finally, we express the expression for the objective function and
its minimization in \eqref{obj_func_min}. Here, we try to minimize the  sum of edge weights (which are
thermal interaction levels lying inside intervals) for edges not belonging to any cluster.
The decision variables $p, s, r$ are the vectors with entries from
$p_{i,c},s_{i,j,c},r_{i,j}$, respectively.
\begin{align} \label{obj_func_min}
 J=\min_{r}\sum_{\substack{(i,j)\in E, \\
 \mu_{i,j}\in I_{i,j}}}(1-r_{i,j})\mu_{i,j}
\end{align}

Letting $p \triangleq [\{p_{i,c}\}] \in \{0,1\}^{N_z \times n}$,
$s \triangleq  [\{s_{i,j,c}\}]\in \mathbb{R}_{+}^{|E|\times n}$,
$r \triangleq  [\{r_{i,j}\}]\in \mathbb{R}_{+}^{|E|}$, and
$\mu \triangleq  [\{\mu_{i,j}\}]\in \mathbb{R}_{+}^{|E|}$, we can write
the optimization problem consisting of the objective function minimization given by \eqref{obj_func_min} and
constraints given \eqref{cfbc_1}- \eqref{cfbc_10} \& \eqref{size_c_1}-\eqref{size_c_2} as
in \eqref{milp_compact_form}. 
\begin{align} \label{milp_compact_form}
& \min_{r}\,(1-r)^T\mu, \quad \mu \in I \subseteq \mathbb{R}_{+}^{|E|}  \nonumber \\
&\hspace{1cm} \text{subject to} \nonumber \\
&\underbrace{\left(
   \begin{array}{ccc}
     A_p^1   & 0        & 0 \\
     0       & A_s^1    & 0 \\
     A_p^2   & A_s^2    & 0 \\
     0       & A_s^3    & A_r^1 \\
     0       & 0        & A_r^2 \\
   \end{array}
 \right)}_{A\in \mathbb{Z}^{d_1\times d_2}}\left(
          \begin{array}{c}
            p \\
            r \\
            s \\
          \end{array}
        \right)
 \le b,\\
& p \in \{0,1\}^{N_z \times n}, \,s \in \mathbb{R}_{+}^{|E|\times n},\,r \in \mathbb{R}_{+}^{|E|}, \nonumber
\end{align}
\noindent where $I$ is the thermal interaction interval vector,  the dimensions of $A$ are
$d_1=2(N_z+n^2)+|E|(3n+5),\, d_2=N_zn+|E|(n+1)$,
and in $A$ the first block row corresponds to \eqref{cfbc_1} \& \eqref{Cluster size constraints},
the second block row to \eqref{cfbc_2}, the third block row to \eqref{cfbc_3}-\eqref{cfbc_5},
the fourth block row to \eqref{cfbc_6}, and the fifth block row to \eqref{cfbc_7}.
 
In the next sections, we will present a stochastic optimization approach and a robust optimization approach
to formulate the uncertainty in the objective function.

\subsection{A stochastic optimization approach to optimal clustering}

In the stochastic optimization framework, the goal is to minimize the expectation of the cost
since the constraints are deterministic and the cost function is the
only part involving uncertainty.
As a result, in \eqref{obj_func_min} we replace all the cost coefficients by their expectation.
To that end, let each interval $I_{ij}$ be divided into $n_d$ sub-intervals
and let $\hat{I}_{ij,k}$ represent the mean of $k$-th sub-interval
with its probability of occurrence $P_{ij,k}$, $k\in \{1, 2, \cdots, n_d \}$.
Defining $\displaystyle\mu_{ij}^s \triangleq \bigg[\Big\{\sum_{k=1}^{n_d}P_{ij,k}\hat{I}_{ij,k}\Big\}\bigg]$, $\mu^s \triangleq  [\{\mu_{i,j}\}] \in \mathbb{R}_{+}^{|E|}$,
the stochastic formulation of the optimization problem becomes
\begin{align} \label{stoc_milp_compact_form}
& \min_{r}\,( 1-r)^T\mu^s \nonumber \\
&\hspace{1cm} \text{subject to} \nonumber \\
&
  A\left(
          \begin{array}{c}
            p \\
            r \\
            s \\
          \end{array}
        \right)
 \le b,\\
& p \in \{0,1\}^{N_z \times n}, \,s \in \mathbb{R}_{+}^{|E|\times n},\,r \in \mathbb{R}_{+}^{|E|}. \nonumber
\end{align}

\subsection{A robust optimization approach to optimal clustering}

In this section, we will present the robust optimization version
of the mixed 0,1 LP problem in \eqref{milp_compact_form}, which has
an uncertain objective function due to the uncertainty in $\mu$.
As a first step, by letting $( 1-r)^T\mu \le z$ and minimizing $z$ (the new objective function
and new variable), we transform the original objective function into a constraint.
Next, defining $\bar{\mu} \triangleq (\mu_{max}+\mu_{min})/2$ and $\mu_{mg}\triangleq \mu_{max}-\bar{\mu}$,
we can write the interval uncertainty vector as
$I=\bar{\mu}+ \text{diag}(\varepsilon_1, \cdots, \varepsilon_{|E|})\mu_{mg}$, where
the entries of the random diagonal matrix satisfy $\big|\varepsilon_i\big| \le 1$.
Following the lines in \cite{Bertsimas_Sim_2004, Li_et_al_2011} for transformation of a robust MILP into an equivalent
deterministic MILP (called the ``robust counterpart"), we obtain the robust counterpart as
in \eqref{robust_milp_compact_form}:

\begin{align} \label{robust_milp_compact_form}
& \hspace{2.4cm}\min_{z}\ z \nonumber  \\
&\hspace{2.25cm} \text{subject to} \nonumber \\
&\underbrace{\left(
   \begin{array}{r:ccc}
 -1   &   0       & -(\bar{\mu}+\mu_{mg})^T        & 0 \\ \hdashline
  0   &   A_p^1   & 0                                     & 0 \\
  0   &   0       & A_s^1                                 & 0 \\
  0   &   A_p^2   & A_s^2                                 & 0 \\
  0   &   0       & A_s^3                                 & A_r^1 \\
  0   &   0       & 0                                     & A_r^2 \\
   \end{array}
 \right)}_{\tilde{A}\in \mathbb{Z}^{(d_1+1)\times (d_2+1)}}\left(
          \begin{array}{c}
            z \\ \hdashline
            p \\
            r \\
            s \\
          \end{array}
        \right)
\\ \nonumber
 &  \hspace{5cm} \le  \underbrace{\left(
       \begin{array}{c}
         (\mu_{mg}-\bar{\mu})^T1 \\ \hdashline
         b \\
       \end{array}
     \right)}_{\tilde{b}}
 ,\\
& z \in \mathbb{R}_{+},\, p \in \{0,1\}^{N_z \times n }, \,s \in \mathbb{R}_{+}^{|E|\times n},\,r \in \mathbb{R}_{+}^{|E|}. \nonumber
\end{align}

\subsection{Definition of performance index and optimality deterrioration/fault propagation metrics}

The performance of any control architecture will be measured through the following performance index (PI):

\begin{equation}
\text{PI}\triangleq e^{-k_u\frac{u_{tot}}{u_{tot-nr}}} \times e^{-k_{y_{v-ave}}\frac{y_{v-ave}}{y_{v-ave-nr}}} \times e^{-k_{y_{v-max}}\frac{y_{v-max}}{y_{v-max-nr}}},
\end{equation}
where the first term takes into account the effect of total consumed energy, the second term
the effect of average comfort temperature violation over all zones and the period over which the system is controlled,
and the final term the effect of maximum comfort temperature violation;
$k_u, k_{y_{v-ave}}, k_{y_{v-max}}$ are weight parameters; $u_{tot-nr}, y_{v-ave-nr}, y_{v-max-nr}$
are parameters used for normalization. Note that PI=1 when $u_{tot}=0$ and no comfort temperature violation,
which is impossible in practice. As a result, PI=1 is an ideal bound, but the closer PI is to 1,
the better the performance of that control architecture is. The only constraint on the weights $k_u, k_{y_{v-ave}}, k_{y_{v-max}}$
is that  PI($1$, no-fault) ($n=1$ corresponds to C-MPC) should have the largest value compared to
all other MPC architectures with/without faults, since fault-free C-MPC is the best solution.

Next, we define two metrics that measure the optimality and sensitivity
to fault propagation of a given control architecture corresponding
to a ``best n-partition ($\text{n}^\star$)".  The first metric is called the ``optimality deterioration metric" (ODM) defined as
\begin{equation} \label{OD_metric}
\text{ODM}(\text{n}^\star, \text{no-fault})\triangleq \frac{\text{PI(1, no-fault)}-\text{PI}(\text{n}^\star, \text{no-fault})}{\text{PI(1, no-fault)}},
\end{equation}
where $\text{PI}(\text{n}^\star, \text{no-fault})$ denotes the performance index
of the control architecture corresponding to the best $n$-partition under no fault.
The second metric called ``fault propagation metric" (FPM) will
be used to quantitatively  determine the sensitivity of an MPC architecture to a
fault in control/measurement equipment in a zone.
This metric is defined as
\begin{equation} \label{FP_metric}
\text{FPM}(\text{n}^\star, \text{fault})\triangleq\frac{\text{PI(1, no-fault)}-\text{PI}(\text{n}^\star, \text{fault})}{\text{PI(1, no-fault)}}.
\end{equation}

Finally, we need to determine which best n-partition is the best partition. To that end,
we define a weighted-performance metric (WPM) as 
\begin{align}
\text{WPM}(\text{n}^\star)\triangleq& \alpha(1-\text{ODM}(\text{n}^\star,\text{no-fault}))+\nonumber \\
                           & (1-\alpha)(1-\text{FPM}(\text{n}^\star,\text{fault})),
\end{align}
where $ 0\le \alpha \le 1$ is a user-parameter. Then, the best n-partition is the one which has
the highest WPM value.

\begin{remark}[Expected performance]
The paradigm of the presented approach is based on the realistic intuition that
the lower the level of thermal interaction between clusters, the more likely higher WPM values
for the decentalizedly controlled clusters. However, there is no guarantee for this.
\end{remark}

\section{Case studies}
\label{sec:Case studies}

In this section we will consider two case studies to demonstrate the  developed
optimal partitioning approach. Due to space limitations,
we will present the results only for the stochastic optimization version. The first case study is a small-scale
multi-zone building, which allows one to analyse all possible connected partitions for  post-assessment (testing
the decentralized controller corresponding to the associated partitioning) and to
determine the success rate of the partitioning method. In both case studies, it was assumed that comfort range is [22, 24] \textdegree{}C and that  each zone temperature can be controlled with a separate heater/cooler. In D-MPC designs,
we had the following assumptions: for a  given cluster, interaction temperatures between zones in that cluster and neighbor zones in other clusters were set to 23 \textdegree{}C (the middle value in the comfort range);
the prediction horizon was taken as 6 hours;
the only constraint  was to keep zone temperatures
in the comfort band whenever possible; the cost function was the 
weighted sum of total energy consumption and
comfort temperature violation where
the weight used for comfort temperature violation penalization
was $10^6$ times the weight used for each control input ($w_{u_i}=1$).
In applying the developed partitioning approach to 
the case studies, we did not use any size and relative size constraints on the formed clusters.

\subsection{Case study 1}

\begin{figure*}[t!]
\centering
\includegraphics[scale=0.275]{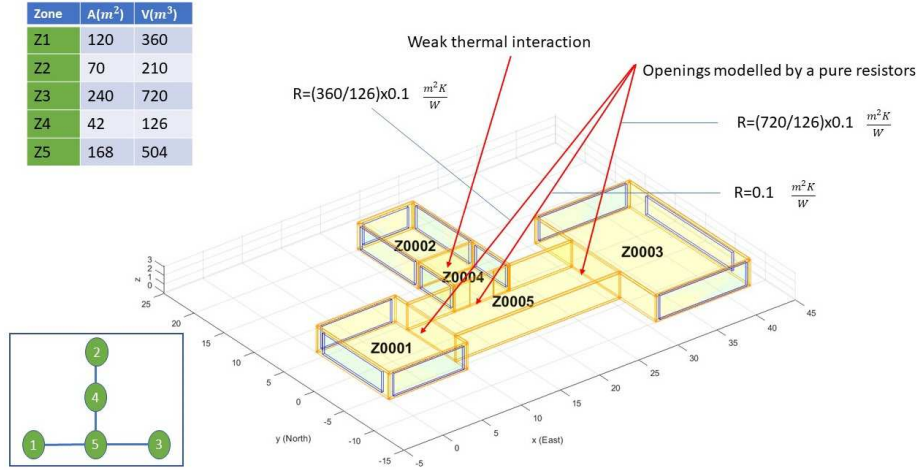}
\caption{A five-zone building, its data and thermal interaction network.}
\label{fig:case_study_1}
\end{figure*}

\begin{figure}[t!]
\centering
\includegraphics[scale=0.375]{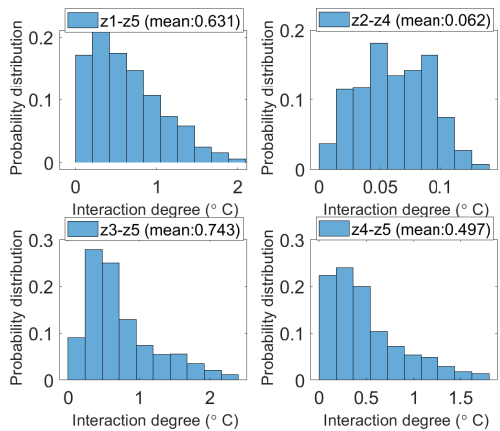}
\caption{Thermal interaction distributions over one year for the first case study.}
\label{fig:case_study_1_ther_inter_dist}
\end{figure}

\begin{figure*}[t!]
\centering
\includegraphics[scale=0.3]{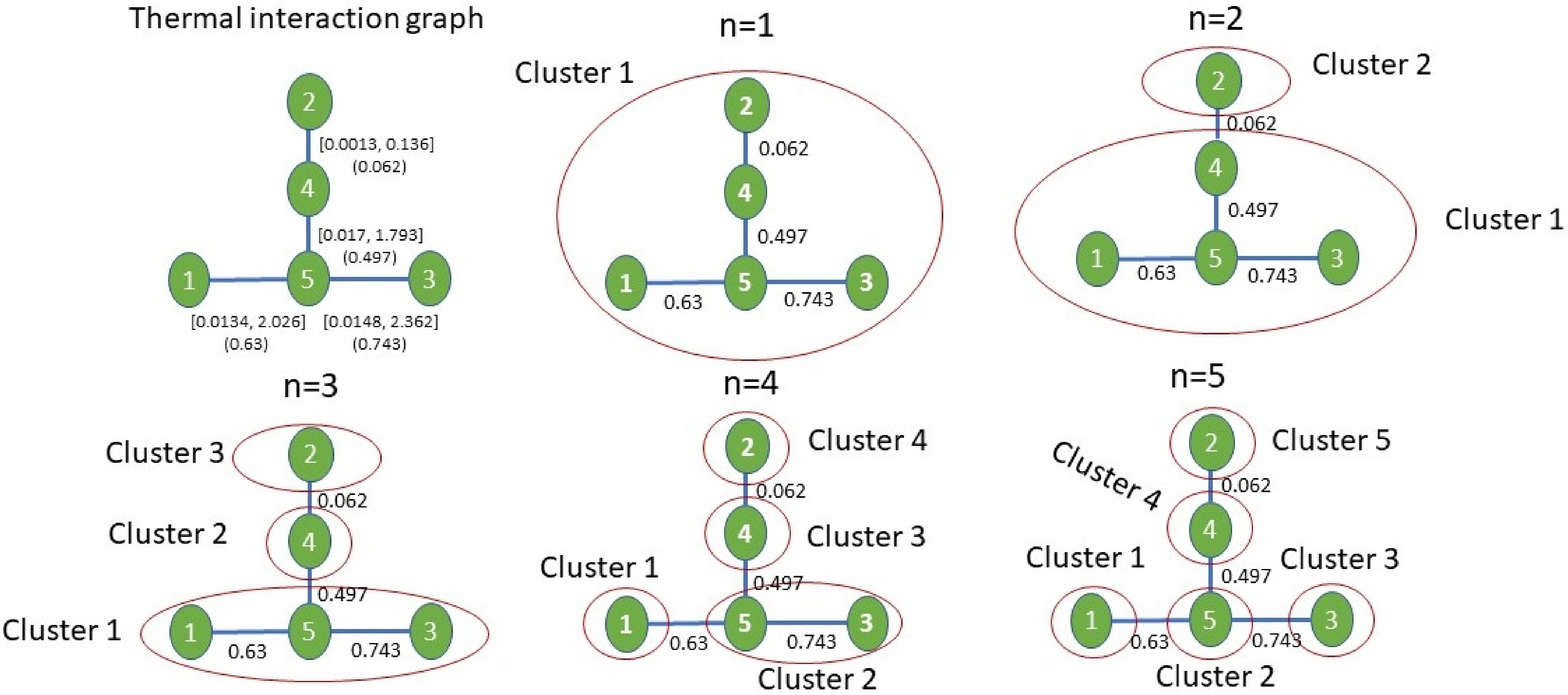}
\caption{Optimal n-partitions for the first case study. Interval and single numbers on edges represent thermal interaction intervals and their mean, respectively.}
\label{fig:optimal_n_partitions.eps}
\end{figure*}

This case study, shown in Figure \ref{fig:case_study_1}, is a 5-zone office building used during 8:00-18:00.
There are significant openings between Z1-Z5, Z3-Z5, and Z4-Z5 which are modeled by pure resistors.
The resistance between Z4-Z5 is taken as $R=0.1 m^2K/W$ and the resistances between  Z1-Z5 and Z3-Z5
are scaled multiples of $R$ where scales are ratios of volumes of Z1 and Z5 to volume of Z4.
The thermal interaction intervals for this case study are:
$\text{I}_\text{1-5}=[0.0134, 2.026]$\textdegree{}C,
$\text{I}_\text{2-4}=[0.0013, 0.136]$\textdegree{}C,
$\text{I}_\text{3-5}=[0.0148, 2.36]$\textdegree{}C,
$\text{I}_\text{4-5}=[0.0171, 1.79]$\textdegree{}C.
The thermal interaction distributions together with their mean values are given in Figure \ref{fig:case_study_1_ther_inter_dist},
which shows that there is no specific pattern.
Optimal n-partitions determined from the developed approach are presented in Figure \ref{fig:optimal_n_partitions.eps}.

To determine whether the optimal n-partitions are really optimal we designed D-MPCs for all possible connected clusters including the optimal n-partitions. For this case study we considered three scenarios: S1-all zone temperatures are controlled,
S2-Z1-Z4 controlled, Z5 uncontrolled (to investigate the thermal effect
of an uncontrolled zone on controlled zones), and S3-there is $\pm$ 10\% error in zone air temperature sensor for all
zones but one zone at a time (to investigate the effect of fault propagation).
The 5-zone building was controlled to keep the zone
temperatures in the comfort band. 
We calculated the average energy per day and per zone ($u_{tot}$),
average temperature violation ($y_{v-ave}$) and maximum temperature violation ($y_{v-max}$).
The results for S1, S2 and S3 for all possible connected partitions are given in the specified columns
in Table \ref{table:case_study1} which also includes the ODM, FPM and WPM
when all zones are controlled. The weight parameters
were taken as $k_u=k_{y_{v-ave}}=k_{y_{v-max}}=1$; $u_{tot-nr}=100 \text{ kWh}, y_{v-ave-nr}=1 \textdegree{C},
y_{v-max-nr}=3 \textdegree{C}$; $\alpha=0.5$.

\begin{table*}[t!]
\center
\caption{Post-evaluation results (over one day) for all connected partitions. The comfort range for controlled zones is [22, 24]\textdegree{}C during 8:00-18:00.}
\label{table:case_study1}
\begin{tabular}{|p{0.1cm}|p{2.15cm}||p{0.85cm}|p{0.82cm}|p{0.8cm}||p{0.8cm}|p{0.8cm}|p{0.8cm}||p{0.85cm}|p{1cm}|p{0.8cm}||p{0.6cm}|p{0.65cm}|p{0.65cm}|}
  \hline
  \multicolumn{2}{|c||}{n and partition} & \multicolumn{3}{c||}{Z1-Z5 controlled} & \multicolumn{3}{c||}{Z1-Z4 controlled} & \multicolumn{3}{c||}{Faulty case} & \multicolumn{3}{c|}{Metrics} \\ \hline
   n & partition &$\text{u}_\text{tot}$ (kWh) & $\text{y}_\text{v-ave}$ (\textdegree{}C) & $\text{y}_\text{v-max}$ (\textdegree{}C) & $\text{u}_\text{tot}$ (kWh) & $\text{y}_\text{v-ave}$ (\textdegree{}C) &$\text{y}_\text{v-max}$ (\textdegree{}C)& $\text{u}_\text{tot}$ (kWh) & $\text{y}_\text{v-ave}$ (\textdegree{}C) & $\text{y}_\text{v-max}$ (\textdegree{}C) & ODM (\%) & FPM (\%) & WPM (\%)\\ \hline \hline
  \rowcolor{green}
  1 & \{1,2,3,4,5\}                 & 58.9649  & 0       & 0        & 49.0952 & 0         & 0        & 58.7341  & 0.1608  & 2.0582 &
  0 &  57.026 &  71.487 \\ \hline
  2 & \{1,2,4,5\},\{3\}             & 58.1501  & 0.0084  & 0.0525   & 48.2976  & 0.0072   & 0.1215   & 58.7303  & 0.1594  & 1.9429 & 1.762 &   55.277 &   71.481 \\ \hline
  \rowcolor{green} 
  2 & \{1,3,4,5\},\{2\}             & 58.9643  & 0.0001  & 0.0004   & 49.0945  & 0.0001   & 0.0004   & 58.6829  & 0.1581  & 1.8642 &   0.016 &   54.007 &   72.989\\ \hline
  2 & \{1,3,5\},\{2,4\}             & 58.8276  & 0.0036  & 0.0382   & 48.9677  & 0.0055   & 0.1101   & 58.6496  & 0.1587  & 1.9275 &  1.488 &   54.978 &   71.767 \\ \hline
  2 & \{1\},\{2,3,4,5\}             & 57.0280  & 0.0276  & 0.1934   & 47.1860  & 0.0293   & 0.5173   & 58.6615  & 0.1645  & 1.8398 &  7.012 &   53.917 &   69.536 \\ \hline
  \rowcolor{green} 
  3 & \{1,3,5\},\{2\},\{4\}         & 58.8270  & 0.0037  & 0.0386   & 48.9672  & 0.0055   & 0.1105   & 58.6340  & 0.1602  & 1.8411 &  1.504 &  53.726 &  72.385\\ \hline
  3 & \{1,4,5\},\{2\},\{3\}         & 58.1495  & 0.0084  & 0.0525   & 48.2969  & 0.0072   & 0.1215   & 58.6336  & 0.1606  & 1.8893 & 1.764 &  54.483 &  71.876 \\ \hline
  3 & \{1,5\},\{2,4\},\{3\}         & 58.0123  & 0.0121  & 0.0630   & 48.1740  & 0.0126   & 0.1215   & 58.6649  & 0.1607  & 1.8526 & 2.325  & 53.942 &  71.866 \\ \hline
  3 & \{1\},\{2,4,5\},\{3\}         & 56.2089  & 0.0361  & 0.1987   & 46.4418  & 0.0359   & 0.5176   & 58.6844  & 0.1666  & 1.8074 & 7.203 &  53.526 &   69.636 \\ \hline
  3 & \{1\},\{2,4\},\{3,5\}         & 56.8894  & 0.0313  & 0.1945   & 47.0676  & 0.0346   & 0.5173   & 58.5631  & 0.1672  & 1.8889 & 
  7.258 &   54.741 &   69.000\\ \hline
  3 & \{1\},\{2\},\{3,4,5\}         & 57.0274  & 0.0276  & 0.1934   & 47.1854  & 0.0293   & 0.5173   & 58.5562  & 0.1678  & 1.7497 &  7.014 &   52.619 &  70.184 \\ \hline
  4 & \{1\},\{2,4\},\{3\},\{5\}     & 56.0699  & 0.0398  & 0.1998   & 46.3273  & 0.0410   & 0.5176   & 58.5528  & 0.1708  & 1.8716 & 7.449 &   54.639  & 68.956 \\ \hline
  \rowcolor{green}
  4 & \{1,5\},\{2\},\{3\},\{4\}     & 58.0118  & 0.0121  & 0.0630   & 48.1734  & 0.0126   & 0.1215   & 58.8271  & 0.1597  & 1.7811 & 
  2.327 &   52.860 &  72.406\\ \hline
  4 & \{1\},\{2\},\{3,5\},\{4\}     & 56.8888  & 0.0313  & 0.1945   & 47.0670  & 0.0346   & 0.5173   & 58.6167  & 0.1687  & 1.8067 &  7.260 &   53.579 &   69.581 \\ \hline
  4 & \{1\},\{2\},\{3\},\{4,5\}     & 56.2083  & 0.0361  & 0.1987   & 46.4412  & 0.0359   & 0.5176   & 58.7196  & 0.1674  & 1.7254 &  7.205 &   52.292 &   70.252\\ \hline
  \rowcolor{green} 5 & \{1\},\{2\},\{3\},\{4\},\{5\} & 56.0693  & 0.0398  & 0.1998   & 46.3267  &  0.0410  & 0.5176   & 58.3571  & 0.1711  & 1.8167 & 7.450 &   53.723 &   69.413 \\
  \hline
\end{tabular}
\end{table*}

\subsection{Case study 2}

\begin{figure*}[t!]
  \subfigure[]{%
    \includegraphics[width=0.45\textwidth]{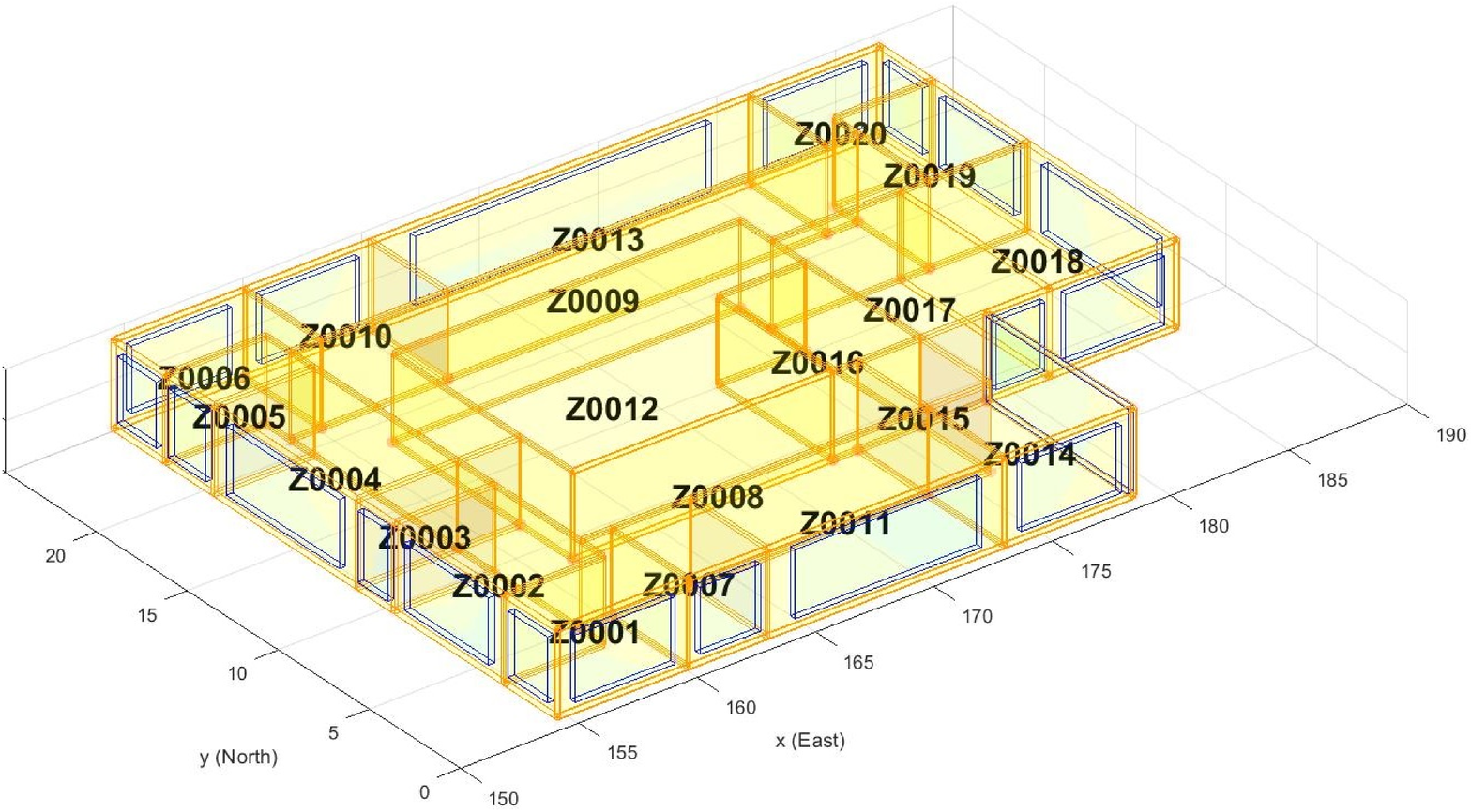} \label{fig:case_study2_building}
  }
  \quad
  \subfigure[]{%
    \includegraphics[width=0.45\textwidth]{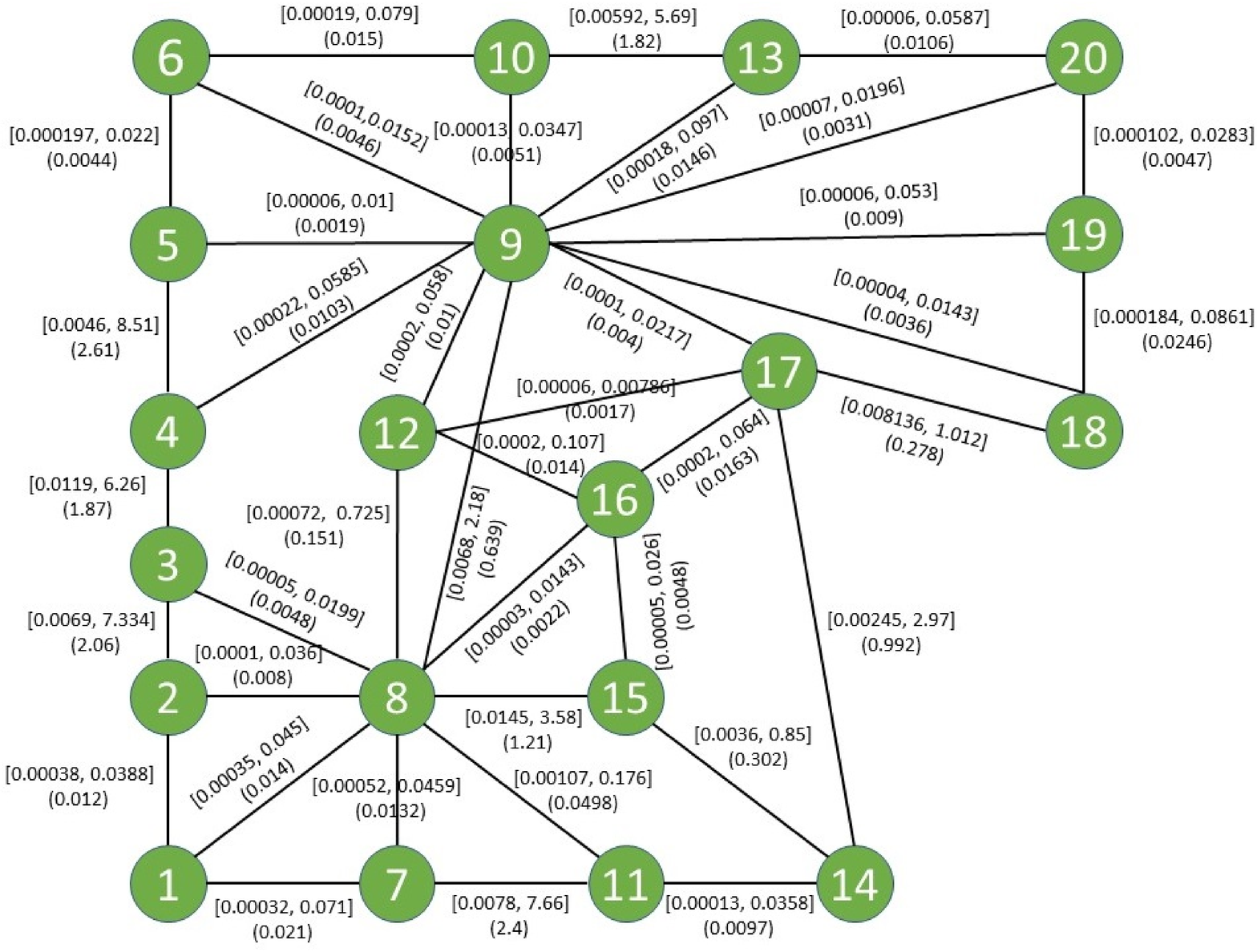} \label{fig:case_study2_ti_graph}
  }
  \caption{(a) A twenty-zone building and (b) its thermal interaction network.
  The values in parenthesis are mean values. Note that zone 9 has 11 thermal interactions.}
  \label{fig:case_study2}
\end{figure*}

The second case study is a 20-zone building with a large number of thermally interacting zones.
The building and its thermal interaction graph are shown in Figure \ref{fig:case_study2}. 
For this case study we considered two scenarios: S1-all zone temperatures are controlled,
and S2-to investigate the effect of fault propagation, we now assumed a different type of a likely fault:  we
assumed that actuators at each zone fail, one at a time, so that no heat/cold is supplied ($u_i$=0). 
The results are given in Table \ref{table:case_study2}. The weight parameters
were taken as $k_u=k_{y_{v-ave}}=k_{y_{v-max}}=1$; $u_{tot-nr}=1000 \text{ kWh}, y_{v-ave-nr}=2 \textdegree{C},
y_{v-max-nr}=5 \textdegree{C}$; $\alpha=0.5$.  

\begin{table*}[t!]
\center
\caption{Post-evaluation results (over one day) for all best partitions. The comfort range for all zones is [22, 24]\textdegree{}C during 8:00-18:00.}
\label{table:case_study2}
\begin{tabular}{|c||l|l|l||l|l|l||l|l|l|}
  \hline
   \multicolumn{1}{|c||}{} & \multicolumn{3}{c||}{Z1-Z20 controlled} & \multicolumn{3}{c||}{Faulty case} & \multicolumn{3}{c|}{Metrics} \\ \hline
  $\text{n}^\star$ &$\text{u}_\text{total}$(kWh)&$\text{y}_\text{v-ave}$ (\textdegree{}C) &$\text{y}_\text{v-max}$(\textdegree{}C)&$\text{u}_\text{total}$(kWh)&$\text{y}_\text{v-ave}$(\textdegree{}C)&$\text{y}_\text{v-max}$(\textdegree{}C) &ODM (\%) & FPM (\%) & WPM (\%)\\ \hline \hline
  $\text{1}^\star$  &795.5808 &0      &0       &823.3647  &0.1104 &3.1123 &0        &50.6112   &74.6944\\ \hline
  $\text{2}^\star$  &795.5657 &0.0001 &0.0006  &823.3480  &0.1105 &3.1125 &0.0127   &50.6133   &74.6870\\ \hline
  $\text{3}^\star$  &795.5608 &0.0001 &0.0007  &823.3427  &0.1105 &3.1125 &0.0143   &50.6136   &74.6861\\ \hline
  $\text{4}^\star$  &795.5601 &0.0001 &0.0007  &823.3417  &0.1105 &3.1125 &0.0155   &50.6139   &74.6853\\ \hline
  $\text{5}^\star$  &795.5499 &0.0001 &0.0017  &823.3302  &0.1105 &3.1128 &0.0346   &50.6168   &74.6743\\ \hline
  $\text{6}^\star$  &795.5481 &0.0001 &0.0017  &823.3282  &0.1105 &3.1128 &0.0345   &50.6167   &74.6744\\ \hline
  $\text{7}^\star$  &795.5144 &0.0002 &0.0017  &823.2911  &0.1106 &3.1128 &0.0342   &50.6172   &74.6743\\ \hline
  $\text{8}^\star$  &795.5106 &0.0002 &0.0018  &823.2866  &0.1106 &3.1129 &0.0378   &50.6178   &74.6722\\ \hline
  $\text{9}^\star$  &795.5081 &0.0002 &0.0018  &823.2839  &0.1106 &3.1129 &0.0379   &50.6178   &74.6721\\ \hline
  $\text{10}^\star$ &795.5057 &0.0002 &0.0018  &823.2815  &0.1106 &3.1129 &0.0380   &50.6179   &74.6721\\ \hline
  $\text{11}^\star$ &795.4738 &0.0003 &0.0030  &823.2476  &0.1107 &3.1132 &0.0621   &50.6218   &74.6581\\ \hline
  $\text{12}^\star$ &730.4916 &0.2251 &2.0331  &756.3917  &0.3330 &3.7592 &36.4961  &58.4863   &52.5088\\ \hline
  $\text{13}^\star$ &709.5264 &0.2912 &2.3103  &733.9544  &0.3987 &3.8677 &40.6440  &59.7976   &49.7792\\ \hline
  $\text{14}^\star$ &686.2700 &0.3751 &3.2300  &710.2009  &0.4793 &4.2173 &51.5292  &63.1284   &42.6712\\ \hline
  $\text{15}^\star$ &683.6270 &0.3855 &3.3964  &707.2868  &0.4904 &4.2870 &53.2361  &63.7347   &41.5146\\ \hline
  $\text{16}^\star$ &641.7458 &0.4901 &3.3964  &663.0952  &0.5903 &4.2768 &53.7213  &63.8693   &41.2047\\ \hline
  $\text{17}^\star$ &610.2526 &0.6849 &3.3965  &630.5260  &0.7778 &4.2813 &56.6738  &66.0434   &38.6414\\ \hline
  $\text{19}^\star$ &601.1079 &0.7466 &3.3965  &621.0750  &0.8379 &4.3016 &57.6042  &66.8703   &37.7627\\ \hline
  $\text{19}^\star$ &571.9436 &0.8562 &3.3966  &590.7952  &0.9422 &4.3017 &58.6784  &67.5875   &36.8670\\ \hline
  $\text{20}^\star$ &550.9360 &1.0015 &4.3932  &569.1118  &1.0813 &4.7148 &67.8492  &71.5529   &30.2989\\
  \hline
\end{tabular}
\end{table*}

\subsection{Discussions of results}
\label{sec:Discussions of results}

The computation time for each partitioning in both case studies is less than one minute using
a laptop with the following hardware specifications: 8GB RAM, Intel(R) Core(TM) i7-8550U CPU  @ 1.80GHz 1.99 GHz.
CPLEX was used as the solver in the solution of the mixed-integer linear programs.  
From Table \ref{table:case_study1} of Case Study 1 we observe that,
when all zone temperatures are controlled, all control architectures (C-MPC and D-MPC) have performances 
which are close to each other on the basis of WPM, even though the building in case study 1 is a highly
thermally-interacting structure to due large openings between zones. 
The explanation for such an observation is the fact that
when all zone temperatures are controlled in a narrow band, then the thermal interaction is not significant,
even though the building has a huge potential for thermal interaction.
When one of the zones was not controlled in Case Study 1,
we observe from Table \ref{table:case_study1} that temperature violation increases in D-MPC cases since thermal interactions increase,
but  the performance of D-MPCs (based on WPM) are still acceptable.
In Table \ref{table:case_study1} green rows indicate the best partition of each n-partition.
When these results are compared with the optimal partitioning results in Figure \ref{fig:optimal_n_partitions.eps},
which are results predicted by our approach, we see that out of five best n-partitions, four of them
are predicted correctly. However, since there is only one possible partition when $n=1,5$, 
it will be more correct to say that out of three best partitions, two of them are predicted correctly by our developed approach
with a success rate of 66.7\%. Based on the WPM, the best control architecture for the
first case study is $n^\star=2$ with the zone partition $\{2\},\{1, 3, 4, 5\}$. 

As regards to the second case study, we see from Table \ref{table:case_study2} that
the best control architecture predicted by our OLBP approach based on WPM is C-MPC. However, note that
the performance of D-MPC with $n^\star=11$ is very close to that of C-MPC,
which again shows that a D-MPC control architecture can work quite well
in control of multi-zone buildings.

\section{Conclusions}
\label{sec:Conclusions}

In this paper, we presented an approach for optimal partitioning of multi-thermal zone buildings for decentralized control.
Both stochastic and robust optimization versions of the developed
approach were presented, and the stochastic version of the algorithm
was demonstrated on two case studies. The first case study was
a small-scale example so that  we were able to obtain all possible partitions
and post-asses the performance of the corresponding controllers.
From the post-assessment, we observed that the success rate of our optimal
partitioning algorithm is $66.7\%$, quite good when the difficulty and complexity
of the optimal partitioning problem are considered. 
Next, we tested the method on a 20-zone building and determined the
optimal partitioning predicted for this building. Moreover, in this paper we presented new metrics (optimality deterioration and 
fault propagation metrics) which can be used in a general decentralized control framework.

The most important findings of this study can be listed as follows: 
(i) the presented  partitioning algorithm can be used effectively
to determine the best partition and the corresponding decentralized control
architecture for multi-zone buildings; (ii) decentralized
control can work very well in office buildings (even for cases where there is a high potential for thermal interaction
between zones) since zone temperatures are, in general, strictly controlled in these buildings during working hours, and as a result, thermal interactions are not significant (except in the morning during the short period
when controllers start to move the uncontrolled zone temperatures to the comfort band);
(iii) since decentralized control can work quite well for office buildings,
for this category of buildings one can use a decentralized control-oriented
modeling approach per zone cluster, which will ease the current bottle-neck (the considerable effort in control model
development \cite{Atam_Helsen_2016}) for the wide-spread application of MPC in office buildings;
(iv) for multi-zone buildings, it is not always correct that  D-MPC will outperform C-MPC
in case of fault propagation: the faults stay more local in D-MPC compared to C-MPC but since the control models
in D-MPC are less accurate (since they are not able to take all thermal interactions into account)
the real performance of D-MPC depends on a combination of these two effects in faulty scenarios. 

A future research direction is to consider a co-design approach where MPC design is explicitly integrated into the partitioning problem. This approach,
although computationally expensive, has a huge
potential to outperform the 
approach presented here.

\balance 

\bibliographystyle{IEEEtran}

\end{document}